\documentclass[10pt,conference,a4paper]{IEEEtran}

\usepackage{epsfig}
\usepackage{psfrag}
\usepackage{amsmath}
\usepackage{amssymb}
\usepackage{amsfonts}
\usepackage{color}
\usepackage{times}

\newcommand{\argmax}{\mathop{\mathrm{argmax}}}
\def\PSNR{\mathrm{ PSNR}}
\def\punit{\, \mathrm}

\begin{document}

\title{Adaptive Joint Spatio-Temporal Error Concealment for Video Communication}
\author{\authorblockN{J\"urgen~Seiler and Andr\'e~Kaup} \vspace{1.6mm}
\authorblockA{\fontsize{10}{10}\selectfont\itshape Chair of Multimedia Communications and Signal Processing, \\University of Erlangen-Nuremberg, Cauerstr. 7, 91058 Erlangen, Germany\\ {\fontsize{9}{9}\selectfont\ttfamily\upshape \{seiler, kaup\}@LNT.de}}
}
\maketitle


\begin{abstract} \label{abstract}
In the past years, video communication has found its application in an increasing number of environments. Unfortunately, some of them are error-prone and the risk of block losses caused by transmission errors is ubiquitous. To reduce the effects of these block losses, a new spatio-temporal error concealment algorithm is presented. The algorithm uses spatial as well as temporal information for extrapolating the signal into the lost areas. The extrapolation is carried out in two steps, first a preliminary temporal extrapolation is performed which then is used to generate a model of the original signal, using the spatial neighborhood of the lost block. By applying the spatial refinement a significantly higher concealment quality can be achieved resulting in a gain of up to 5.2 dB in PSNR compared to the unrefined underlying pure temporal extrapolation.
\end{abstract}


\section{Introduction} \label{sec:introduction}

In the past years the transmission of video signals over wireless channels or over IP-based networks became more and more common. Unfortunately, these environments are very error-prone and single bit errors, as well as complete packet losses may occur. In order to be able to effectively transmit in these cases anyway, modern video codecs as e.\ g.\  the H.264/AVC use two strategies to cope with these challenges, according to \cite{Stockhammer2005}. The first strategy is the error resilience that aims at protecting the bitstream against transmission errors. If this strategy fails and transmission errors occur, the second strategy, the error concealment has to be applied. Even though error concealment strategies are not part of the actual standard, they are very important for displaying a pleasant video signal despite the errors and for reducing distortion introduced by error propagation. 

In the scope of this contribution we will focus on the second strategy, the concealment of errors. A good survey of this topic can be found in \cite{Wang1998}. According to them, the concealment of lost picture blocks is performed by extrapolating the signal from correctly received areas either in spatial or in temporal direction. For the former one, only information from the spatial neighborhood of the lost block is used for extrapolating the signal and therewith concealing the loss. On the other hand, the temporal error concealment algorithms only use information from temporally adjacent frames to extrapolate the signal into the area of the lost block. Normally, this is obtained by estimating the motion of a sequence and replacing the lost block with one from another frame that is shifted to compensate the estimated motion. Due to the high temporal correlation between temporally adjacent frames, temporal extrapolation algorithms mostly are superior to spatial ones according to \cite{Bopardikar2005}. Unfortunately, both groups of algorithms suffer from the circumstance that they either use only temporal or only spatial information for the extrapolation and ignore the other one.

To cope with this shortcoming, we propose a new extrapolation algorithm, the adaptive joint spatio-temporal extrapolation and its application to error concealment. There, a preliminary temporal extrapolation is performed which afterwards is refined using information from the spatial neighborhood. Thus the temporal extrapolated block better fits into the erroneous frame. As temporal as well as spatial information is used for the signal extrapolation, visual as well as objective concealment abilities can be improved significantly compared to pure temporal algorithms.


\section{Spatio-temporal extrapolation} \label{sec:st-extrapolation}

\begin{figure}
 \centering
 	\psfrag{m}[t][t][0.8]{$m$}%
	\psfrag{n}[t][t][0.8]{$n$}%
	\psfrag{x}[t][t][0.8]{$x$}%
	\psfrag{x0}[t][t][0.8]{$x_0$}%
	\psfrag{y}[t][t][0.8]{$y$}%
	\psfrag{y0}[t][t][0.8]{$y_0$}%
	\psfrag{t}[t][t][0.8]{$t$}%
	\psfrag{t1}[t][t][0.8]{$t=\tau-1$}%
	\psfrag{t0}[t][t][0.8]{$t=\tau$}%
	\psfrag{Block}[l][l][0.8]{$\mathcal{B}$}%
	\psfrag{Rgb}[l][l][0.8]{$\mathcal{R}$}%
	\psfrag{Pgb}[l][l][0.8]{$\mathcal{P}=\mathcal{R}\cup\mathcal{B}$}%
 \includegraphics[width=0.45\textwidth]{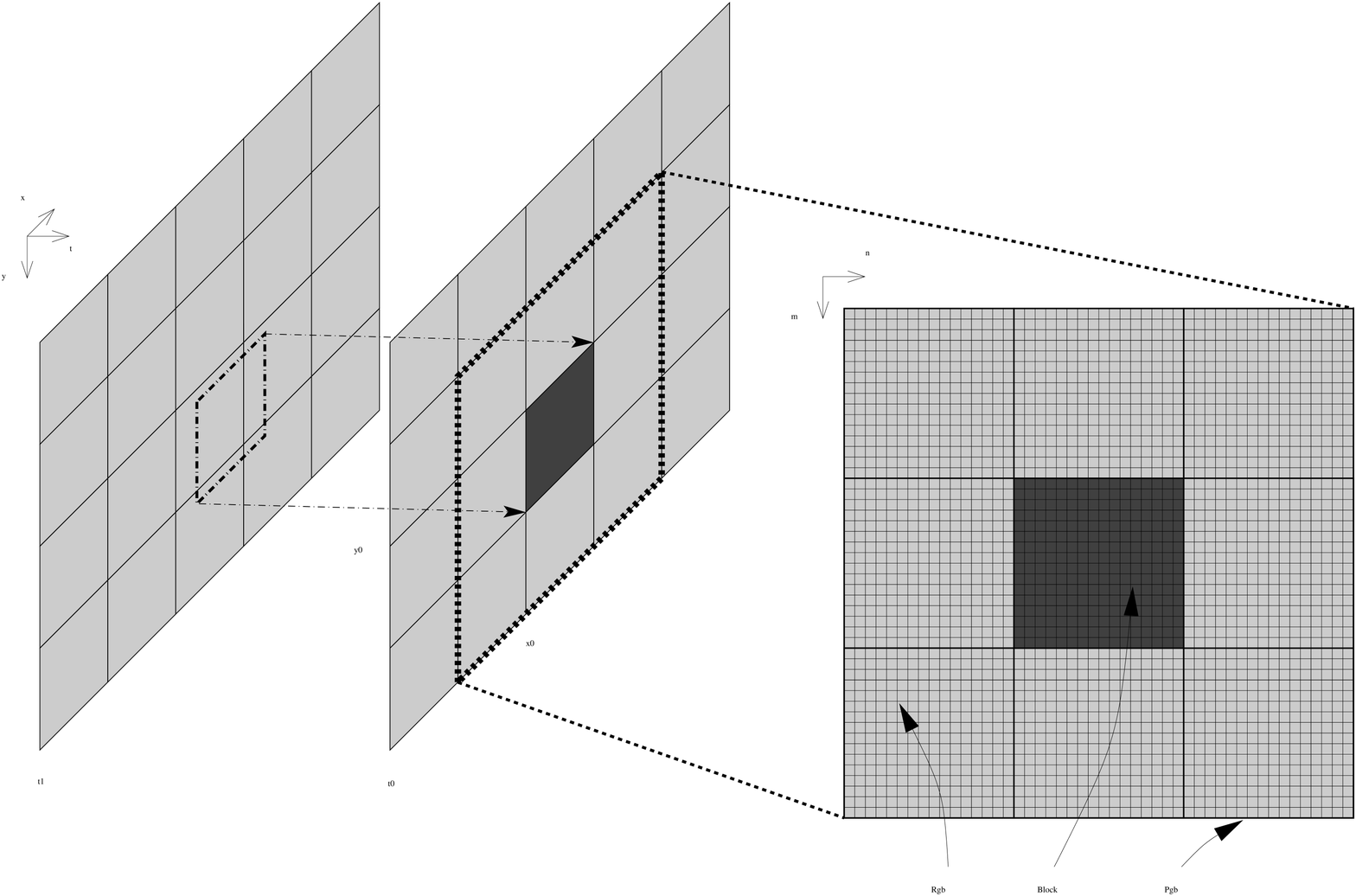}
 \caption{Left: two frames of the video sequence $v\left[x,y,t\right]$ with an isolated block loss at $\left(x_0,y_0\right)$. Right: Projection area $\mathcal{P}$ consisting of the preliminarily temporal extrapolated block $\mathcal{B}$ and correctly received adjacent blocks $\mathcal{R}$}\vspace{-0.4cm}
 \label{fig:extrapolation_area}
\end{figure}

In the following, the spatio-temporal extrapolation is outlined for the example of concealment of isolated block losses. Block losses are chosen for presentational reasons and the algorithm can easily be applied to other extrapolation scenarios as well. The regarded video sequence is described by a three-dimensional signal $v\left[x,y,t\right]$ with the two spatial coordinates $x$ and $y$ and the temporal coordinate $t$. This sequence is corrupted by an isolated block loss of size $B\times B$ samples with the top left corner at $\left(x_0,y_0\right)$. As mentioned above, the adaptive joint spatio-temporal extrapolation operates in two steps. In the first step, a preliminary temporal extrapolation is performed based on previous correctly received frames. This step is described more detailed in the next subsection. After having determined a temporal estimate for the lost block, a model of the signal is generated for incorporating spatial information. Therefore a new signal $f\left[m,n\right]$ is generated consisting of the preliminarily temporal extrapolated block, depicted by $\mathcal{B}$ and the surrounding correctly received blocks, subsumed in area $\mathcal{R}$. Areas $\mathcal{B}$ and $\mathcal{R}$ together form the projection area $\mathcal{P}$ of size $M\times N$ samples. The projection area is depicted by a new coordinate system with the two spatial coordinates $m$ and $n$. Fig.\ \ref{fig:extrapolation_area} is used to illustrate the relationship between the video sequence $v\left[x,y,t\right]$ and the projection area $\mathcal{P}$. The model generation is performed on area $\mathcal{P}$, and after having finished the model generation the samples corresponding to area $\mathcal{B}$ are cut out of the model and are used to conceal the lost block. 

\subsection{Preliminary temporal extrapolation}

The first step of the spatio-temporal extrapolation is a preliminary temporal extrapolation. Therefore, a block from a previous frame has to be determined that is used to replace the lost block. The simplest temporal extrapolation method is the Temporal Replacement (TR). There, the lost block is replaced by the block located at the same spatial position in the previous frame. Although this method provides acceptable results for static and very slow sequences, it fails in case of motion. A more sophisticated method for temporal extrapolation is the Extended Boundary Matching Algorithm (EBMA) from \cite{Lam1993}. Instead of using the block located at the same spatial position for concealing the block loss, it seeks for the block in the previous frame that minimizes the boundary error to the correctly received adjacent blocks. Thus, EBMA is able to compensate the motion in the sequence and to provide a decent visual concealment, as the temporally extrapolated block fits to the adjacent blocks. The Decoder Motion Vector Estimation (DMVE) from \cite{Zhang2000a} uses a different criterion for determining which block to use from the previous frame to conceal the lost block. DMVE takes an area around the lost block and compares this area with shifted ones from the previous frame. Afterwards the shift is chosen as estimate for the motion vector that minimizes the error between the area in the actual frame and the corresponding one in the previous frame. Then, the lost block is replaced by the block from the previous frame that is shifted according to the estimated motion vector. Although there are many other techniques for concealing lost blocks by temporal extrapolation, within the scope of this contribution we will focus on the three just mentioned algorithms for performing the preliminary temporal extrapolation, as they are widely used and often referenced. 

\begin{figure}
 \centering
 	\psfrag{x}[t][t][0.8]{$x$}%
	\psfrag{x0}[t][t][0.8]{$x_0$}%
	\psfrag{xd}[t][t][0.8]{$x_0+x_\mathrm{d}$}%
	\psfrag{y}[t][t][0.8]{$y$}%
	\psfrag{y0}[t][t][0.8]{$y_0$}%
	\psfrag{yd}[t][t][0.8]{$y_0+y_\mathrm{d}$}%
	\psfrag{t}[t][t][0.8]{$t$}%
	\psfrag{t1}[t][t][0.8]{$t=\tau-1$}%
	\psfrag{t0}[t][t][0.8]{$t=\tau$}%
	\psfrag{D}[t][t][0.8]{$\mathcal{D}$}
	\psfrag{Dd}[t][t][0.8]{$\mathcal{D}_\mathrm{d}$}
 \includegraphics[width=0.3\textwidth]{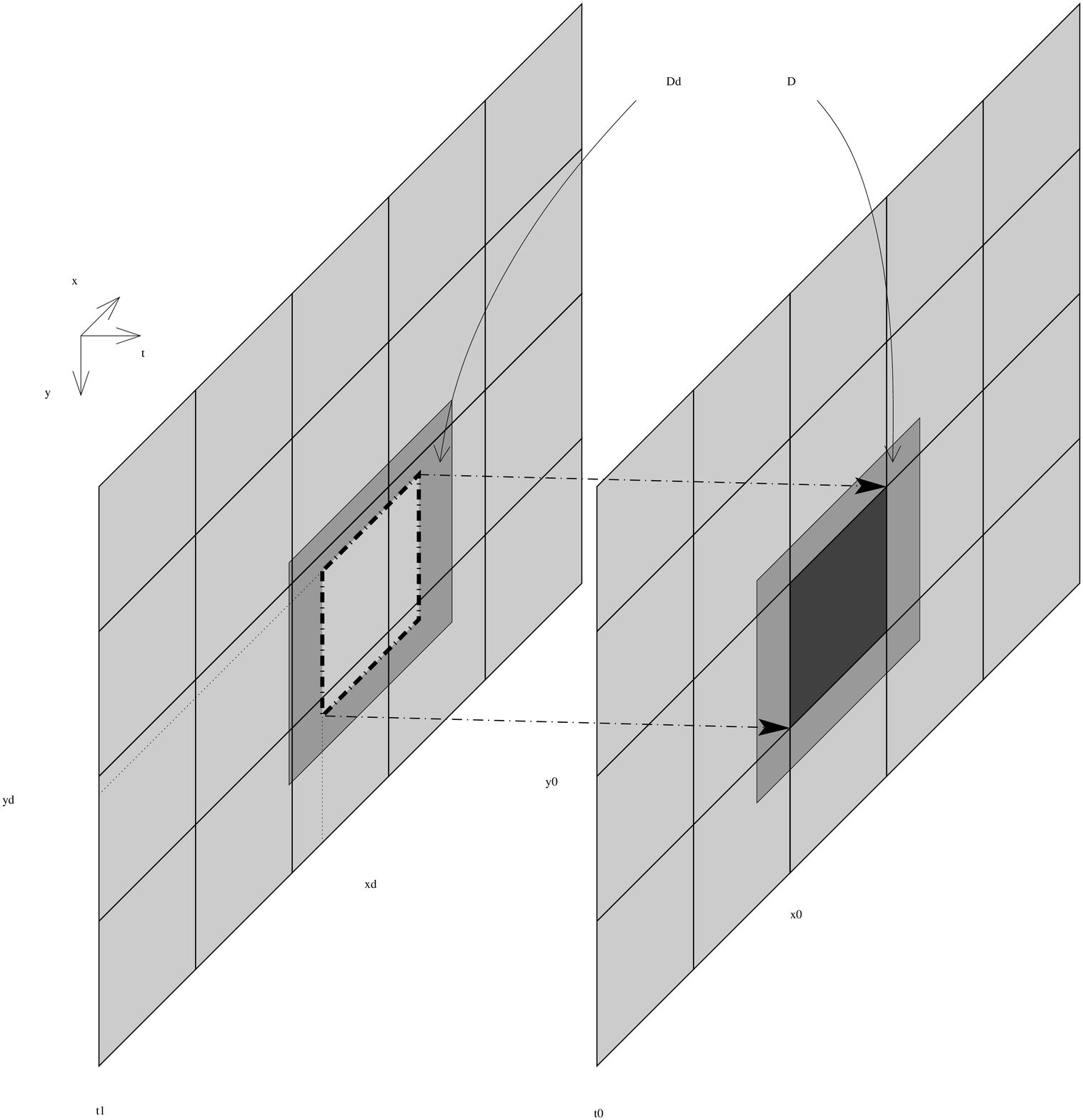}
 \caption{Area $\mathcal{D}$ and displaced area $\mathcal{D}_\mathrm{d}$ for evaluating the extrapolation quality}\vspace{-0.4cm}
 \label{fig:border_area}
\end{figure}

For the subsequent spatial refinement the quality of the temporal extrapolation has to be evaluated in order to determine the relevance of the temporal extrapolation for the model generation. Therefore the border $\mathcal{D}$ around the lost block, but covered by correctly received samples, is examined. As described before, the preliminary temporal extrapolation is obtained by replacing the lost block at $\left(x_0,y_0\right)$ in frame $t=\tau$ with a block from frame $t=\tau-1$ at $\left(x_0+x_\mathrm{d},y_0+y_\mathrm{d}\right)$ with the displacement vector $\left(x_\mathrm{d}, y_\mathrm{d}\right)$. So the displaced area $\mathcal{D}_\mathrm{d}$ in frame $t=\tau-1$ can be used to estimate the extrapolation quality similarly to the error criterion in DMVE \cite{Zhang2000a}. Fig.\ \ref{fig:border_area} illustrates the relationship between $\mathcal{D}$ and $\mathcal{D}_\mathrm{d}$. The comparison of the signal in area $\mathcal{D}$ and $\mathcal{D}_\mathrm{d}$ leads to the estimated temporal extrapolation error
\begin{equation}
 \hat{e}_{\mathrm{t}} = \sqrt{\frac{1}{\left|\mathcal{D}\right|} \sum_{\left(x,y\right) \in \mathcal{D}} \left(v\left[x,y,\tau\right] - v\left[x+x_\mathrm{d},y+y_\mathrm{d}, \tau-1\right] \right)^2}
\label{eq:etee}
\end{equation}
with $\left|\mathcal{D}\right|$ denoting the cardinality of $\mathcal{D}$.  

\subsection{Spatial refinement}

After having determined the preliminary temporal extrapolation, the projection area $\mathcal{P}$ is set up. Therefore, the preliminary estimated block is put in the center of the area, surrounded by adjacent correctly received blocks, as illustrated in Fig.\ \ref{fig:extrapolation_area}. This union is described by the signal $f\left[m,n\right]$. The spatial refinement is obtained by performing a two-dimensional approximation of the signal in order to generate a model of the signal in whole area $\mathcal{P}$. The model generation is based on the selective approximation from \cite{Kaup1998} and \cite{Seiler2008} and aims at combining the spatial information from the surrounding blocks with the pure temporally extrapolated block. For this, the signal $f\left[m,n\right]$ is approximated by the parametric model 
\begin{equation}
 g\left[m,n\right] = \sum_{\forall k \in \mathcal{K}} c_k \varphi_k \left[m,n\right]
\end{equation}
which is a weighted superposition of mutually orthogonal basis functions $\varphi_k \left[m,n\right]$. In general, every set of mutually orthogonal two-dimensional basis functions could be used, but for incorporating the spatial information well, the functions have to be spatially delocalized. The set of all basis functions used for the model generation is subsumed in $\mathcal{K}$ and the weighting factors $c_k$ are denoted as expansion coefficients.

\begin{figure}
 \centering
\psfrag{x05}[t][t][1.0]{$10$}%
\psfrag{x06}[t][t][1.0]{$20$}%
\psfrag{x07}[t][t][1.0]{$30$}%
\psfrag{x08}[t][t][1.0]{$40$}%
\psfrag{v05}[r][r][1.0]{$10$}%
\psfrag{v06}[r][r][1.0]{$20$}%
\psfrag{v07}[r][r][1.0]{$30$}%
\psfrag{v08}[r][r][1.0]{$40$}%
\psfrag{z05}[r][r][1.0]{$0$}%
\psfrag{z06}[r][r][1.0]{$0.05$}%
\psfrag{z07}[r][r][1.0]{$0.1$}%
\psfrag{z08}[r][r][1.0]{$0.15$}%
 \includegraphics[width=0.4\textwidth]{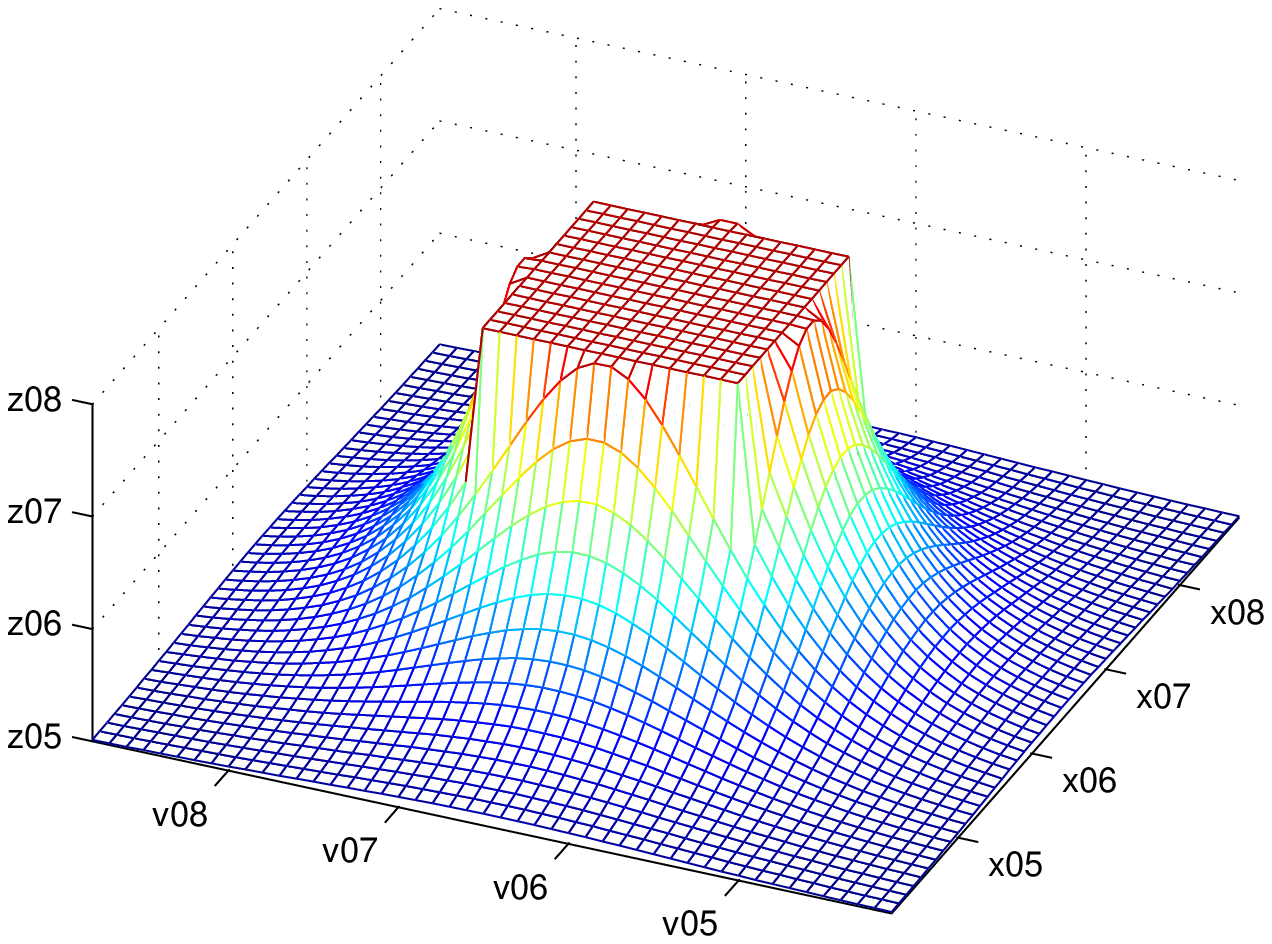}
 \caption{Example weighting function $w\left[m,n\right]$ with $\hat{\rho}=0.8$ and $\mu=0.16$}\vspace{-0.4cm}
 \label{fig:weighting-function}
\end{figure}

According to \cite{Kaup1998} and \cite{Seiler2008}, the parametric model is generated iteratively starting with
\begin{equation}
 g^{\left(0\right)} \left[m,n\right] = 0
\end{equation}
and the approximation error signal $r \left[m,n\right]$ before the first iteration step
\begin{equation}
 r^{\left(0\right)} \left[m,n\right] = f \left[m,n\right]-g^{\left(0\right)}\left[m,n\right] = f\left[m,n\right].
\end{equation} 
In every iteration step one basis function is added to the parametric model. The added basis function and its corresponding expansion coefficient are denoted by index $u$. Resulting, in the $\nu$-th iteration step the parametric model and the approximation error are updated according to
\begin{equation}
 g^{\left(\nu\right)} \left[m,n\right] = g^{\left(\nu-1\right)} \left[m,n\right] + c_u^{\left(\nu\right)} \varphi_u\left[m,n\right]
\end{equation}
and
\begin{equation}
 r^{\left(\nu\right)} \left[m,n\right] = r^{\left(\nu-1\right)} \left[m,n\right] - c_u^{\left(\nu\right)} \varphi_u\left[m,n\right]
\end{equation}
In order to determine the best basis function to be added to the parametric model in the $\nu$-th iteration the weighted projection of $r^{\left(\nu-1\right)}\left[m,n\right]$ on all basis functions is examined, resulting in the projection coefficients
\begin{equation}
 p_k^{\left(\nu \right)} = \frac{\displaystyle \sum_{\left(m,n\right) \in \mathcal{P}} r^{\left(\nu-1\right)} \left[m,n\right] \cdot \varphi_k \left[m,n\right] \cdot w \left[m,n\right]}{\displaystyle \sum_{\left(m,n\right) \in \mathcal{P}} w \left[m,n\right] \cdot \varphi_k^2 \left[m,n\right] }
\end{equation}
for all basis functions. Thereby 
\begin{equation}
 w \left[m,n\right] = \left\{ \begin{array}{ll} \mu &,\ \forall \left(m,n\right) \in \mathcal{B} \\ \rho\left[m,n\right] &,\ \forall \left(m,n\right) \in \mathcal{R} \end{array} \right. 
\end{equation}
denotes the weighting function that controls the influence pixels have on the approximation process depending on their position. Thus, all pixels belonging to the preliminary temporally extrapolated block get the same weight for the model generation, whereas the weight of the surrounding samples depends on their position relative to the testblock. This is obtained by weighting the surrounding pixels according to an isotropic model  for exponentially reducing the weight with an increasing distance to the lost block. The isotropic model is described by 
\begin{equation}
\rho\left[m,n\right] = \hat{\rho}^{\sqrt{ \left(m-\frac{M-1}{2}\right)^2 + \left(n-\frac{N-1}{2}\right)^2}}
\end{equation}
with $\hat{\rho}$ controlling the decay. In Fig.\ \ref{fig:weighting-function} an example weighting function is shown. There, the temporally extrapolated block is weighted by $\mu=0.16$ and the surrounding blocks with the decay parameter $\hat{\rho}=0.8$.

Since the temporal extrapolation quality varies depending on the regarded block, the weight $\mu$ for area $\mathcal{B}$ is adapted according to the estimated temporal extrapolation error $\hat{e}_{\mathrm{t}}$ from (\ref{eq:etee}). In the case that $\hat{e}_{\mathrm{t}}$ exceeds a given maximum extrapolation error $e_\mathrm{max}$, the temporal extrapolation can be regarded as unreliable and the weight $\mu$ is set to $0$. This yields a pure spatial intra frame extrapolation and corresponds to the Frequency Selective Extrapolation proposed in \cite{Seiler2008}. But if $\hat{e}_{\mathrm{t}}$ tends toward $0$, indicating that the temporal extrapolation already is very good, area $\mathcal{B}$ has to get a high weight for the model generation. Thus, $\mu$ is set to the maximum value of the isotropic model, equal to $\hat{\rho}^{B/2}$. If $\hat{e}_{\mathrm{t}}$ is between $0$ and $e_{\mathrm{max}}$, \mbox{$\mu$ decreases} linearly between the maximum value and $0$ which can be expressed by  
\begin{equation}
\mu = \left\{ \begin{array}{lll} \hat{\rho}^{B/2}\cdot\left(1-\frac{\hat{e}_\mathrm{t}}{e_\mathrm{max}}\right) & \mbox{for} & 0<\hat{e}_\mathrm{t}<e_\mathrm{max} \\ 0 & \mbox{else} & \end{array} \right.
\end{equation}

After all projection coefficients have been determined, from all possible basis functions the one is chosen that minimizes the distance between the approximation error signal $r^{\left(\nu-1\right)}\left[m,n\right]$ and the weighted projection onto the corresponding basis function. This is also the basis function that maximizes the approximation error energy's decrement. The index $u$ of this basis function is determined according to 
\begin{equation}
 u = \argmax_{k=0, ...,\left|\mathcal{P}\right|-1} \left(p_k^{\left(\nu \right)^2} \cdot \sum_{\left(m,n\right)\in \mathcal{P}} w\left[m,n\right]\cdot \varphi^2_k\left[m,n\right]\right).
\end{equation}
In order to determine an estimate $\hat{c}_u^{\left(\nu\right)}$ for the real expansion coefficient from the projection coefficient $p_u^{\left(\nu\right)}$ the orthogonality deficiency caused by the weighting function has to be compensated. Although the basis function are mutually orthogonal with respect to the projection area $\mathcal{P}$ they are not orthogonal anymore when evaluated in combination with the weighting function. They are still close to orthogonality, so this circumstance is called orthogonality deficiency. The orthogonality deficiency causes that the projection coefficient $p_u^{\left(\nu\right)}$ incorporates portions from basis functions unlike $\varphi_u\left[m,n\right]$ as well. For obtaining a good estimate $\hat{c}_u^{\left(\nu\right)}$ for the actually present expansion coefficient, the fast orthogonality deficiency compensation proposed in \cite{Seiler2008} is used, leading to
\begin{equation}
 \hat{c}_u^{\left(\nu\right)} = \gamma \cdot p_u^{\left(\nu\right)} \ , \ \gamma \in ]0,1] .
\end{equation}
The factor $\gamma$ is constant and from the range between $0$ and $1$. It is used to avoid running the risk that a basis function is overemphasized in the model due to orthogonality deficiency, disturbing the further model generation. If in an iteration step the estimated expansion coefficient $\hat{c}_u^{\left(\nu\right)}$ is chosen smaller than the real one, the remaining part will be added to the model in a later iteration. 

Finally at the end of each iteration step the index of the chosen basis function is added to the set $\mathcal{K}$ of all used basis functions, if not already done. Additionally, the corresponding expansion coefficient is updated.
\begin{equation}
 \mathcal{K} = \mathcal{K} \cup u \hspace{0.75cm} \mbox{and} \hspace{0.75cm} \hat{c}_u = \hat{c}_u + \hat{c}_u^{\left(\nu\right)}
\end{equation}

After finishing the iterations for generating the parametric model, area $\mathcal{B}$ is cut out from $g\left[m,n\right]$ and is used for concealing the lost block.

\begin{figure}
 \centering
 \psfrag{Lost block}[c][c][0.8]{Lost block}
 \psfrag{Preliminary temporal}[c][c][0.8]{Preliminary temporal}
 \psfrag{extrapolation}[c][c][0.8]{extrapolation}
 \psfrag{Evaluation of the}[c][c][0.8]{Evaluation of the}
 \psfrag{temporal extrapolation}[c][c][0.8]{temporal extrapolation}
 \psfrag{Weight calculation for}[c][c][0.8]{Weight calculation for}
 \psfrag{temporally extrapolated block}[c][c][0.8]{temporally extrapolated block}
 \psfrag{Iterative model generation}[c][c][0.8]{Iterative model generation}
 \psfrag{by selective approximation}[c][c][0.8]{by selective approximation}
 \psfrag{Replacement of lost block}[c][c][0.8]{Replacement of lost block}
 \psfrag{by area B from model gmn}[c][c][0.8]{by area $\mathcal{B}$ from model $g\left[m,n\right]$}
 \includegraphics[width=0.27\textwidth]{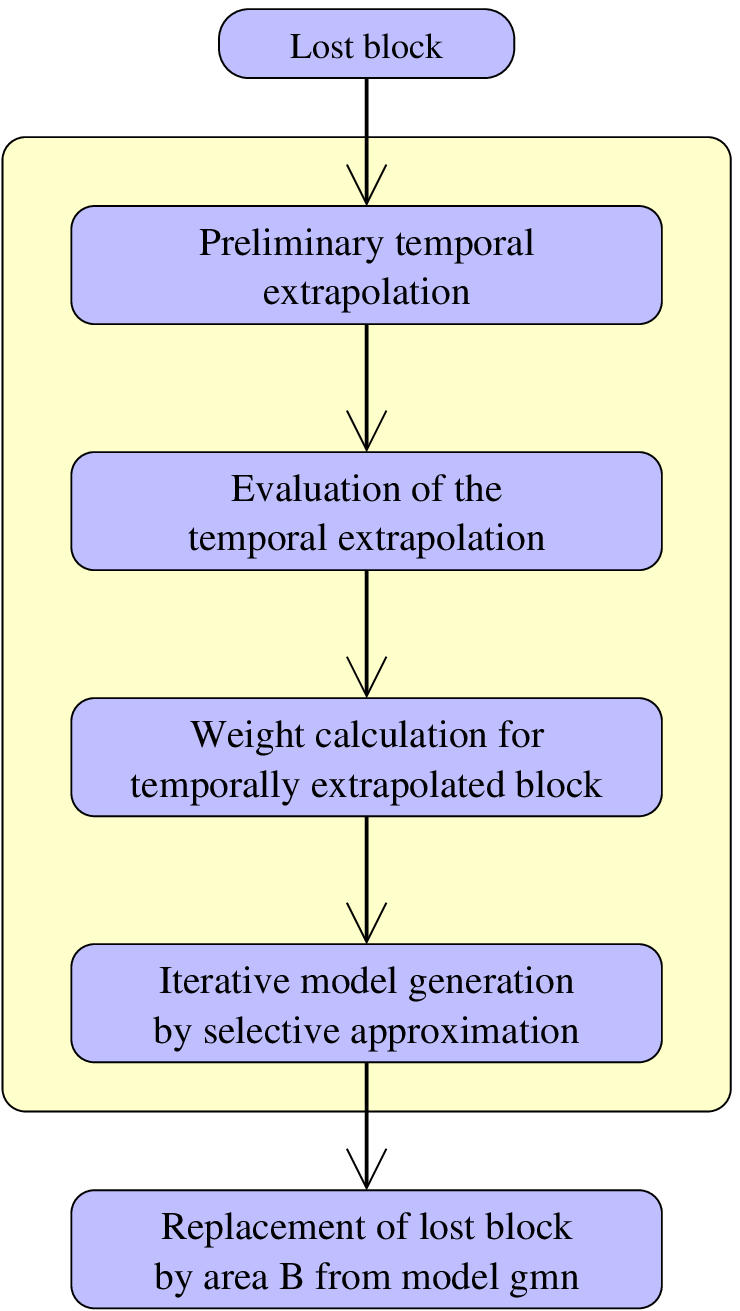}
 \caption{Block diagram for adaptive joint spatio-temporal error concealment}\vspace{-0.4cm}
 \label{fig:bsb}
\end{figure}

The block diagram in Fig.\ \ref{fig:bsb} illustrates the main steps of the of the previously described spatio-temporal extrapolation.

\section{Simulation setup and results}\label{sec:results}

\begin{table*}
\centering
\caption{Concealment results for direct and refined temporal extrapolation in terms of $\PSNR$}
\label{tab:results}
\begin{tabular}{|c||c|c|c||c|c|c||c|c|c|}
\hline & \multicolumn{3}{|c||}{TR} & \multicolumn{3}{|c||}{DMVE \cite{Zhang2000a}} & \multicolumn{3}{|c|}{EBMA \cite{Lam1993}}\\
\hline & direct & refined & gain & direct & refined & gain & direct &refined &gain\\
\hline\hline ``City'' & $26.9\punit{dB}$ & $27.5\punit{dB}$ & $0.6 \punit{dB}$ & $28.9\punit{dB}$ & $28.9\punit{dB}$ & $0.0 \punit{dB}$ & $22.2\punit{dB}$ & $23.4\punit{dB}$ & $1.2 \punit{dB}$\\
\hline ``Crew'' & $28.5\punit{dB}$ & $30.9\punit{dB}$& $2.4 \punit{dB}$ & $30.6\punit{dB}$ & $33.0\punit{dB}$ & $2.4 \punit{dB}$& $28.9\punit{dB}$ & $30.7\punit{dB}$& $1.8 \punit{dB}$\\
\hline ``Discovery City'' & $19.8\punit{dB}$ & $26.6\punit{dB}$ & $6.8 \punit{dB}$& $22.2\punit{dB}$ & $27.4\punit{dB}$ & $5.2 \punit{dB}$& $22.5\punit{dB}$ & $27.1\punit{dB}$ & $4.6 \punit{dB}$\\
\hline ``Foreman'' & $26.4\punit{dB}$ & $28.2\punit{dB}$ & $1.8 \punit{dB}$& $32.9\punit{dB}$ & $33.6\punit{dB}$ & $0.7 \punit{dB}$& $29.7\punit{dB}$ & $30.8\punit{dB}$& $1.1 \punit{dB}$\\
\hline ``Vimto'' & $22.1\punit{dB}$ & $25.7\punit{dB}$ & $3.6 \punit{dB}$& $26.2\punit{dB}$ & $28.7\punit{dB}$ & $2.5 \punit{dB}$& $26.5\punit{dB}$ & $28.8\punit{dB}$& $2.3 \punit{dB}$\\
\hline
\end{tabular} \vspace{-0.35cm}
\end{table*}

For demonstrating the abilities of the adaptive joint spatio-temporal extrapolation the concealment of block losses in the CIF sequences ``City'', ``Crew'', ``Discovery City'' ``Foreman'', and ``Vimto'' is evaluated. Therefore, in every frame from number $4$ to $150$, $25$ blocks of size $16\times 16$ pixels are cut out. The luminance component of these lost blocks is concealed by means of the spatially refined temporal extrapolation and is compared to the original blocks in terms of $\PSNR$. The basis functions used for generating the model are the functions of the two-dimensional discrete Fourier transform as for these basis functions an efficient implementation in the transform domain exists, according to \cite{Kaup2005}. Additionally, \cite{Seiler2008, Kaup2005} show that this set of basis functions is especially suitable for reconstructing smooth areas, as well as edges and noise like areas. As the area $\mathcal{R}$ of the correctly received blocks consists of eight blocks surrounding the lost block, the complete projection area $\mathcal{P}$ is of size $48\times48$ samples. For generating the parametric model, $200$ iterations are performed with the parameter of the weighting function $\hat{\rho}$ chosen to $0.8$ and the orthogonality deficiency compensation factor $\gamma$ to $0.75$. The temporal extrapolation quality is evaluated by means of the border $\mathcal{D}$ of $8$ pixels width and the maximum temporal extrapolation error $e_\mathrm{max}$ is set to $25$. 

\begin{figure}
 \centering
\psfrag{s01}[t][t]{\color[rgb]{0,0,0}\setlength{\tabcolsep}{0pt}\begin{tabular}{c}Frame\end{tabular}}%
\psfrag{s02}[b][b]{\color[rgb]{0,0,0}\setlength{\tabcolsep}{0pt}\begin{tabular}{c}$\PSNR$\end{tabular}}%
\psfrag{s03}[b][b]{\color[rgb]{0,0,0}\setlength{\tabcolsep}{0pt}\begin{tabular}{c}\end{tabular}}%
\psfrag{s06}[][]{\color[rgb]{0,0,0}\setlength{\tabcolsep}{0pt}\begin{tabular}{c} \end{tabular}}%
\psfrag{s07}[][]{\color[rgb]{0,0,0}\setlength{\tabcolsep}{0pt}\begin{tabular}{c} \end{tabular}}%
\psfrag{s08}[l][l][0.95]{\color[rgb]{0,0,0}DMVE refined}%
\psfrag{s17}[l][l][0.95]{\color[rgb]{0,0,0}DMVE}%
\psfrag{s18}[l][l][0.95]{\color[rgb]{0,0,0}DMVE refined}%
\psfrag{x12}[t][t][1.]{$0$}%
\psfrag{x13}[t][t][1.]{$50$}%
\psfrag{x14}[t][t][1.]{$100$}%
\psfrag{x15}[t][t][1.]{$150$}%
\psfrag{v12}[r][r][1.]{$0$}%
\psfrag{v13}[r][r][1.]{$5$}%
\psfrag{v14}[r][r][1.]{$10$}%
\psfrag{v15}[r][r][1.]{$15$}%
\psfrag{v16}[r][r][1.]{$20$}%
\psfrag{v17}[r][r][1.]{$25$}%
\psfrag{v18}[r][r][1.]{$30$}%
\psfrag{v19}[r][r][1.]{$35$}%
\psfrag{v20}[r][r][1.]{$40$}%
\vspace{-0.3cm} \includegraphics[width=0.43\textwidth]{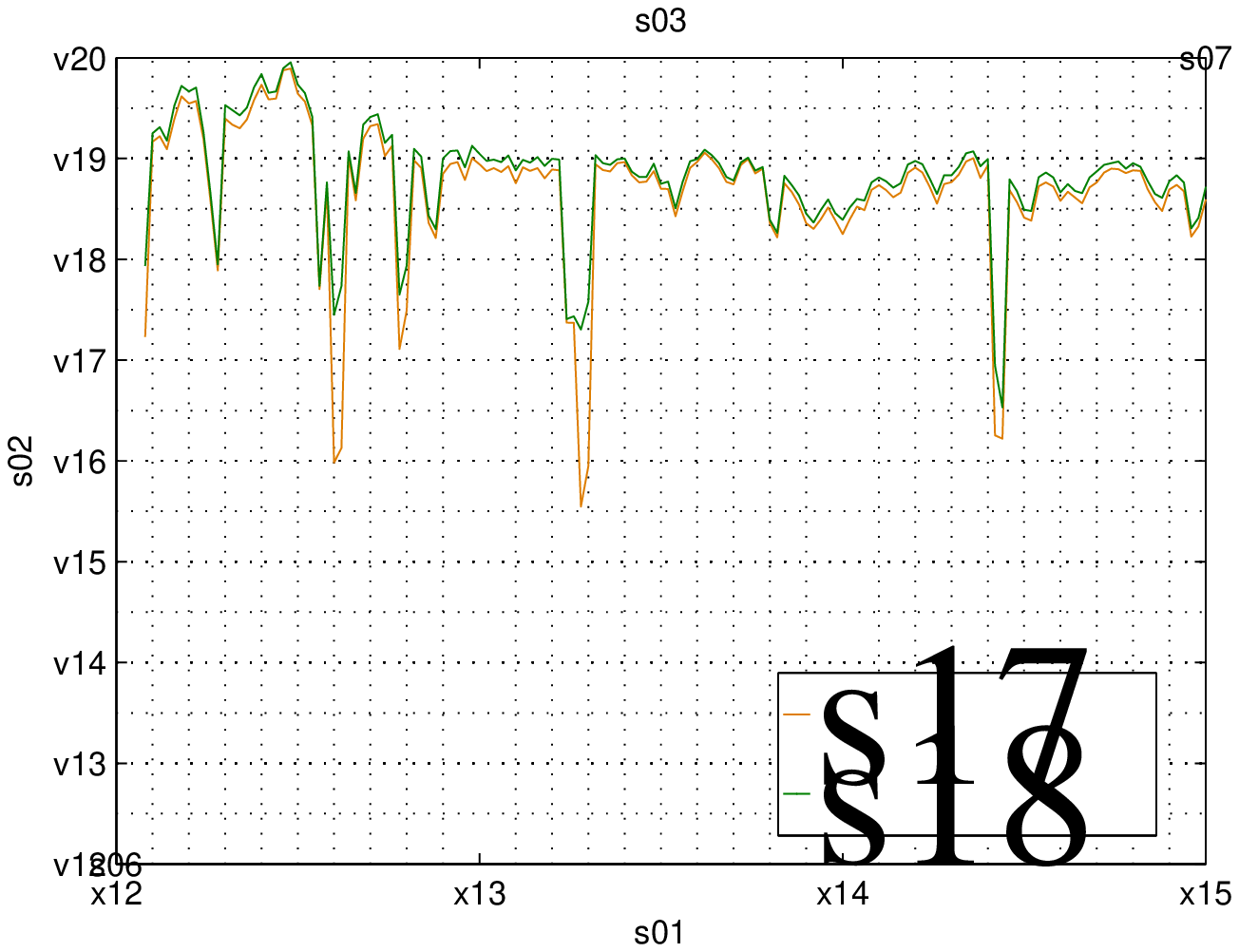}
\caption{$\PSNR$ per frame for test sequence ``Crew'' with error concealment by pure DMVE and DMVE with spatial refinement}\vspace{-0.4cm}
\label{fig:psnr_crew}
\end{figure}

\begin{figure}
 \centering
\psfrag{s01}[t][t]{\color[rgb]{0,0,0}\setlength{\tabcolsep}{0pt}\begin{tabular}{c}Frame\end{tabular}}%
\psfrag{s02}[b][b]{\color[rgb]{0,0,0}\setlength{\tabcolsep}{0pt}\begin{tabular}{c}$\PSNR$\end{tabular}}%
\psfrag{s03}[b][b]{\color[rgb]{0,0,0}\setlength{\tabcolsep}{0pt}\begin{tabular}{c}\end{tabular}}%
\psfrag{s06}[][]{\color[rgb]{0,0,0}\setlength{\tabcolsep}{0pt}\begin{tabular}{c} \end{tabular}}%
\psfrag{s07}[][]{\color[rgb]{0,0,0}\setlength{\tabcolsep}{0pt}\begin{tabular}{c} \end{tabular}}%
\psfrag{s08}[l][l][0.95]{\color[rgb]{0,0,0}DMVE refined}%
\psfrag{s13}[l][l][0.95]{\color[rgb]{0,0,0}DMVE}%
\psfrag{s14}[l][l][0.95]{\color[rgb]{0,0,0}DMVE refined}%
\psfrag{x12}[t][t][1.]{$0$}%
\psfrag{x13}[t][t][1.]{$50$}%
\psfrag{x14}[t][t][1.]{$100$}%
\psfrag{x15}[t][t][1.]{$150$}%
\psfrag{v12}[r][r][1.]{$0$}%
\psfrag{v13}[r][r][1.]{$5$}%
\psfrag{v14}[r][r][1.]{$10$}%
\psfrag{v15}[r][r][1.]{$15$}%
\psfrag{v16}[r][r][1.]{$20$}%
\psfrag{v17}[r][r][1.]{$25$}%
\psfrag{v18}[r][r][1.]{$30$}%
\psfrag{v19}[r][r][1.]{$35$}%
\psfrag{v20}[r][r][1.]{$40$}%
\psfrag{v21}[r][r][1.]{$45$}%
\psfrag{v22}[r][r][1.]{$50$}%
\includegraphics[width=0.43\textwidth]{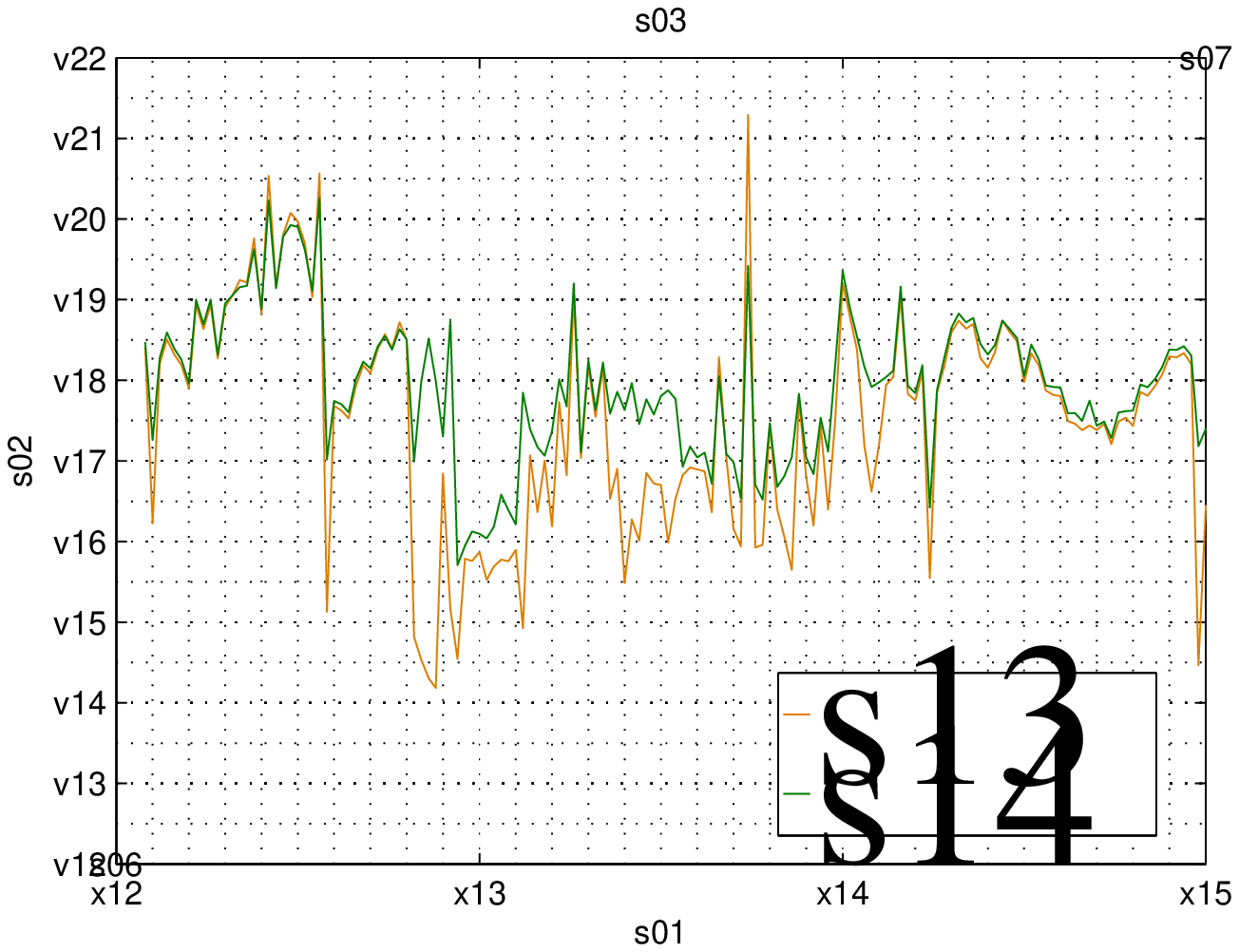}
\caption{$\PSNR$ per frame for test sequence ``Discovery City'' with error concealment by pure DMVE and DMVE with spatial refinement}\vspace{-0.4cm}
\label{fig:psnr_discovery_city}
\end{figure}

\begin{figure}
 \centering
\psfrag{s01}[t][t]{\color[rgb]{0,0,0}\setlength{\tabcolsep}{0pt}\begin{tabular}{c}Frame\end{tabular}}%
\psfrag{s02}[b][b]{\color[rgb]{0,0,0}\setlength{\tabcolsep}{0pt}\begin{tabular}{c}$\PSNR$\end{tabular}}%
\psfrag{s03}[b][b]{\color[rgb]{0,0,0}\setlength{\tabcolsep}{0pt}\begin{tabular}{c}\end{tabular}}%
\psfrag{s06}[][]{\color[rgb]{0,0,0}\setlength{\tabcolsep}{0pt}\begin{tabular}{c} \end{tabular}}%
\psfrag{s07}[][]{\color[rgb]{0,0,0}\setlength{\tabcolsep}{0pt}\begin{tabular}{c} \end{tabular}}%
\psfrag{s08}[l][l][0.95]{\color[rgb]{0,0,0}DMVE refined}%
\psfrag{s13}[l][l][0.95]{\color[rgb]{0,0,0}DMVE}%
\psfrag{s14}[l][l][0.95]{\color[rgb]{0,0,0}DMVE refined}%
\psfrag{x12}[t][t][1.]{$0$}%
\psfrag{x13}[t][t][1.]{$50$}%
\psfrag{x14}[t][t][1.]{$100$}%
\psfrag{x15}[t][t][1.]{$150$}%
\psfrag{v12}[r][r][1.]{$0$}%
\psfrag{v13}[r][r][1.]{$5$}%
\psfrag{v14}[r][r][1.]{$10$}%
\psfrag{v15}[r][r][1.]{$15$}%
\psfrag{v16}[r][r][1.]{$20$}%
\psfrag{v17}[r][r][1.]{$25$}%
\psfrag{v18}[r][r][1.]{$30$}%
\psfrag{v19}[r][r][1.]{$35$}%
\psfrag{v20}[r][r][1.]{$40$}%
\vspace{-0.2cm} \includegraphics[width=0.43\textwidth]{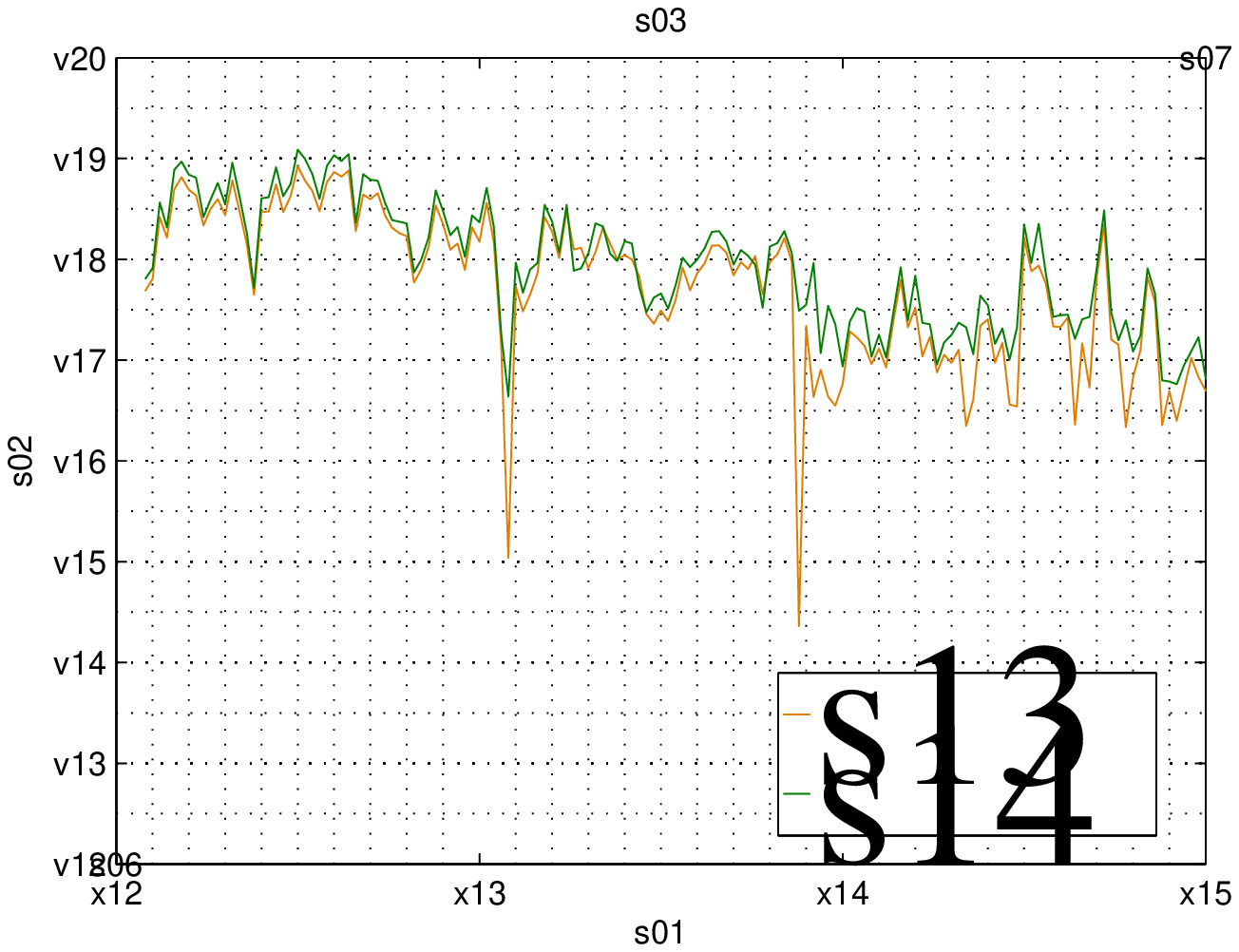}
\caption{$\PSNR$ per frame for test sequence ``Vimto'' with error concealment by pure DMVE and DMVE with spatial refinement}\vspace{-0.4cm}
\label{fig:psnr_vimto}
\end{figure}

As mentioned before, the temporal extrapolation is performed either by Temporal Replacement (TR), or Extended Boundary Matching Algorithm (EBMA), or Decoder Motion Vector Estimation (DMVE). In Table \ref{tab:results} the concealment quality is listed in terms of $\PSNR$ for these three algorithms, on the one hand by directly using them and on the other hand by applying the proposed refinement step on top of the temporal extrapolation. Thereby the usage of Temporal Replacement for the underlying temporal extrapolation is to illustrate, that the spatial refinement is able to perform well, even if the preliminary extrapolation only is poor. For a realistic comparison to unrefined temporal extrapolation, DMVE and BMA are better suited. Regarding all sequences, the $\PSNR$ can be increased by the refinement, whereas the magnitude of the gain depends on the sequence. Considering sequences where the temporal extrapolation already performs well, as e.\ g.\ ``City'' the spatial refinement leads to small gains. But for sequences as e.\ g.\ ``Discovery City'' or ``Vimto'' where the temporal extrapolation often fails, a remarkable increment of more than $5 \punit{dB}$ in terms of $\PSNR$ can be obtained. Regarding the underlying temporal extrapolation, it can be seen that the absolute extrapolation abilities of the spatial refinement depend on the preliminary temporal extrapolation. But independently of the temporal extrapolation, the extrapolation quality can be improved compared to the pure temporal extrapolation.  Fig.\ \ref{fig:psnr_crew}, Fig.\ \ref{fig:psnr_discovery_city}, and Fig.\ \ref{fig:psnr_vimto} show the $\PSNR$ per frame for the sequences ``Crew'', ``Discovery City'', and ``Vimto''. The plots illustrate that for most of the frames the refinement leads to a slight increment in $\PSNR$ but in the case that the temporal extrapolation fails the refinement still reaches a high extrapolation quality, producing the large overall gains. The drops in $\PSNR$ for the pure temporal extrapolation are either produced by scene changes or by singular blocks that cannot be extrapolated well only in temporal direction. Thus, the gain in $\PSNR$ produced by the spatially refined is achieved in two ways. On the one hand, temporally already well extrapolated blocks are improved by incorporating the spatial surrounding. On the other hand, in the case that the temporal extrapolation does not provide good results, implicitly more weight is put on the spatial extrapolation, ending in a pure spatial extrapolation if the spatial extrapolation completely fails.

\begin{figure*}
 \centering
 \includegraphics[width=0.98\textwidth]{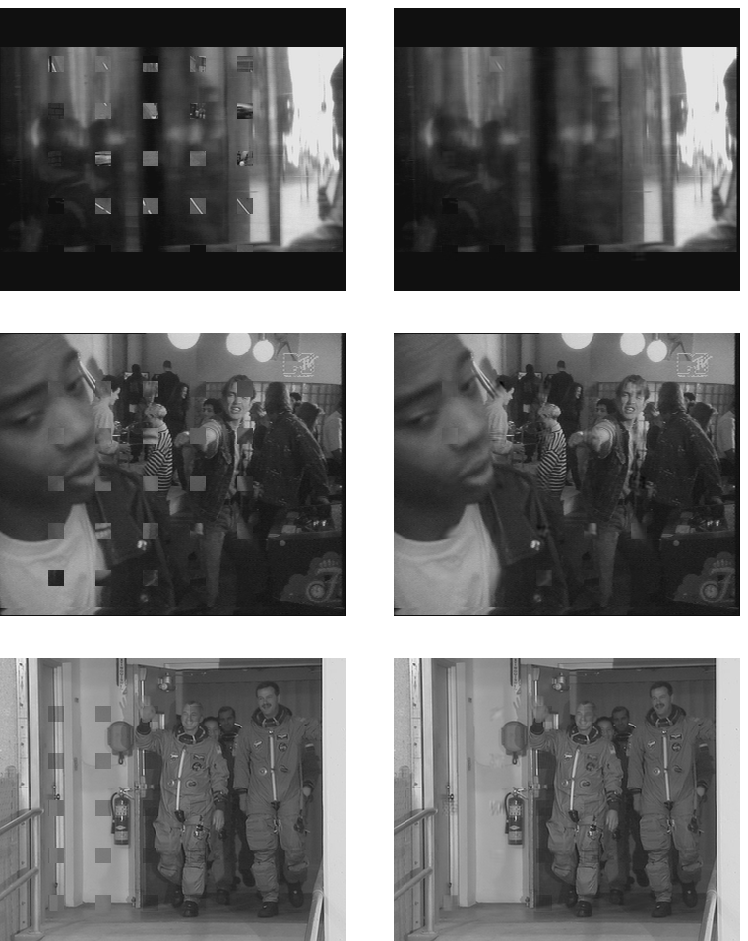}
 \caption{Visual error concealment results for temporally poorly extrapolated example frames. Left column: Error concealment by DMVE. Right column: Error concealment by DMVE with spatial refinement. Top: Frame $56$ from ``Discovery City''. Mid: Frame $54$ from ``Vimto''. Bottom: Frame $30$ from ``Crew'' }
 \label{fig:visual_results}
\end{figure*}

Fig.\ \ref{fig:visual_results} shows some visual results for concealment of isolated block losses that cannot be extrapolated well in temporal direction only due to high motion, scene changes or flash lights. On the left side, for sequences ``Discovery City'', ``Vimto'', and ``Crew'' the lost blocks are concealed by DMVE, on the right side, the spatial refinement step is applied to the preliminary temporal extrapolation performed by DMVE. Here, the concealment only in temporal direction fails due to scene changes in the sequences ``Discovery City'' and ``Vimto'' and due to the flash lights in the sequence ``Crew''. Comparing the unrefined and refined extrapolated blocks, the abilities of the spatial refinement become apparent, resulting in very good visual extrapolation results. 


\section{Conclusion} \label{sec:conclusion}

Within the scope of this contribution we introduced the adaptive joint spatio-temporal extrapolation with its application to error concealment.  The proposed algorithm is an efficient way of combining temporal and spatial information to form an enhanced extrapolation of the signal. Additionally, the idea of the algorithm to use several sources of information in order to form a model of the signal is a very generic one and can be applied to other signal extrapolation tasks as well. For the task of error concealment the proposed spatially refined temporal extrapolation leads to significantly improved signal extrapolation, resulting in more than $5 \punit{dB}$ $\PSNR$ gain and a better visual concealment of the lost blocks. Nevertheless, further investigation will focus on using subpixel accuracy for the preliminary temporal extrapolation in order to improve the initial as well as final extrapolation quality.

\begin{thebibliography}{1}
	\providecommand{\url}[1]{#1}
	\def\UrlFont{\rmfamily}
	\providecommand{\newblock}{\relax}
	\providecommand{\bibinfo}[2]{#2}
	\providecommand\BIBentrySTDinterwordspacing{\spaceskip=0pt\relax}
	\providecommand\BIBentryALTinterwordstretchfactor{4}
	\providecommand\BIBentryALTinterwordspacing{\spaceskip=\fontdimen2\font plus
		\BIBentryALTinterwordstretchfactor\fontdimen3\font minus
		\fontdimen4\font\relax}
	\providecommand\BIBforeignlanguage[2]{{%
			\expandafter\ifx\csname l@#1\endcsname\relax
			\typeout{** WARNING: IEEEtran.bst: No hyphenation pattern has been}%
			\typeout{** loaded for the language `#1'. Using the pattern for}%
			\typeout{** the default language instead.}%
			\else
			\language=\csname l@#1\endcsname
			\fi
			#2}}
	
	\bibitem{Stockhammer2005}
	T.~Stockhammer and M.~H. Hannuksela, ``{H}.264/{AVC} video for wireless
	transmission,'' \emph{IEEE Wireless Communications}, vol.~12, no.~4, pp.
	6--13, Aug. 2005.
	
	\bibitem{Wang1998}
	Y.~Wang and Q.-F. Zhu, ``Error control and concealment for video communication:
	a review,'' \emph{Proceedings of the IEEE}, vol.~86, no.~5, pp. 974--977, May
	1998.
	
	\bibitem{Bopardikar2005}
	A.~S. Bopardikar, O.~I. Hillestad, and A.~Perkis, ``Temporal concealment of
	packet-loss related distortions in video based on structural alignment,'' in
	\emph{Proceedings of the Eurescom summit 2005}, Heidelberg, Germany, April
	2005.
	
	\bibitem{Lam1993}
	W.-M. Lam, A.~R. Reibman, and B.~Liu, ``Recovery of lost or erroneously
	received motion vectors,'' in \emph{Proc. Int. Conf. on Acoustics, Speech,
		and Signal Processing (ICASSP)}, April 1993, pp.
	417--420.
	
	\bibitem{Zhang2000a}
	J.~Zhang, J.~F. Arnold, and M.~F. Frater, ``A cell-loss concealment technique
	for {MPEG}-2 coded video,'' \emph{IEEE Trans. Circuits Syst. Video Technol.},
	vol.~10, no.~4, pp. 659--665, June 2000.
	
	\bibitem{Kaup1998}
	A.~Kaup and T.~Aach, ``Coding of segmented images using shape-independent basis
	functions,'' \emph{IEEE Trans. Image Process.}, vol.~7, no.~7, pp. 937--947,
	July 1998.
	
	\bibitem{Seiler2008}
	J.~Seiler and A.~Kaup, ``Fast orthogonality deficiency compensation for
	improved frequency selective image extrapolation,'' in \emph{Proc. Int. Conf.
		on Acoustics, Speech, and Signal Processing (ICASSP)}, Las Vegas, USA, 31.
	March - 4. April 2008, pp. 781--784.
	
	\bibitem{Kaup2005}
	A.~Kaup, K.~Meisinger, and T.~Aach, ``Frequency selective signal extrapolation
	with applications to error concealment in image communication,'' \emph{{I}nt.
		{J}. {E}lectron. {C}ommun. ({AE}{\"U})}, vol.~59, pp. 147--156, June 2005.
	
\end{thebibliography}
\end{document}